\documentstyle[12pt]{article}
\def\d{\partial}

\def\g{\gamma}

\def\P{\phi}

\def\s{\sigma}

\def\e{\epsilon}

\def\is{\equiv}

\def\pmb#1{\setbox0=\hbox{#1}%
\kern.0em\copy0\kern-\wd0
\kern-.04em\copy0\kern-\wd0
\kern.08em\copy0\kern-\wd0
\kern-.04em\raise.0433em\box0 } 	%poor man's bold macro (Tex Book)

\def\half{{\textstyle{1 \over 2}}}

\def\bop#1{\setbox0=\hbox{$#1M$}\mkern1.5mu
        \vbox{\hrule height0pt depth.04\ht0
        \hbox{\vrule width.04\ht0 height.9\ht0 \kern.9\ht0
        \vrule width.04\ht0}\hrule height.04\ht0}\mkern1.5mu}

\begin{document}

\newcommand{\inv}[1]{{#1}^{-1}} %inverse

\renewcommand{\theequation}{\thesection.\arabic{equation}}
\newcommand{\beq}{\begin{equation}}
\newcommand{\eeq}[1]{\label{#1}\end{equation}}
\newcommand{\ber}{\begin{eqnarray}}
\newcommand{\eer}[1]{\label{#1}\end{eqnarray}}
%\begin{titlepage}
\begin{center}
        December, 1996		\hfill    OSLO-TP 20-96\\
			        \hfill    USITP-96-18\\
				\hfill    hep-th/9612213\\

\vskip .1in

{\large \bf Classical Interactions for Tensionless Strings}
\vskip .2in

{\bf Bj\o rn Jensen} \footnotemark \\

\footnotetext{e-mail address: bjensen@gluon.uio.no}
\bigskip

{\em Institute of Physics \\
University of Oslo\\
Box 275,\\
N-0314 Blindern Oslo NORWAY}\\

\vskip .15in

{\bf Ulf Lindstr\"om} \footnotemark \\

\footnotetext{e-mail address: ul@vana.physto.se, ulfl@boson.uio.no}
\bigskip

\begin{minipage}[t]{2in}
 {\begin{center}{\em  Institute of Theoretical Physics \\
University of Stockholm \\
Box 6730 \\
S-113 85 Stockholm SWEDEN}\\
\end{center}}
\end{minipage} \ and \ 
\begin{minipage}[t]{2in}
 {\begin{center}{\em Institute of Physics \\
University of Oslo\\
Box 275,\\
N-0314 Blindern Oslo NORWAY}\\
\end{center}}
\end{minipage}

\vskip .1in
\end{center}
\vskip .4in
\begin{center} {\bf ABSTRACT } \end{center}
\begin{quotation}\noindent
Using an ``action at a distance'' formulation we probe the possible classical
interactions for tensionless strings, (the $T\rightarrow 0$ limit
of the ordinary bosonic string.)
 We find $G_{\mu\nu}$ and $B_{\mu\nu}$ type
interactions but no dilaton interactions.
\end{quotation}
\vfill
\eject

\normalsize
\section{Introduction}

In this brief note we apply the idea of direct string-string
interactions to tensionless strings. This approach was used for
particle interactions by Feynman and Wheeler \cite{Feynman}, and for
strings by Kalb and Ramond \cite{Kalb}.

The tensionless strings discussed in this note are the $T\to 0$ limit
of ordinary (bosonic) strings, with essentially the same relation to
tensile strings as massless particles have to massive ones. Their
relation to the tensionless ``non critical'' $6D$ strings recently
discussed in the context of $M$-theory, \cite {witt1}-\cite{sixten}, 
is difficult to judge
since the dynamics of the $6D$-strings is not yet known. The fact that
we find possible classical graviton interactions, and that the
non-critical strings should not couple to gravitons, speaks against
such a relation.

In the search for a relativistic theory of gravitation, it was early
realized that gravitation cannot be mediated by a scalar field only. A
heuristic way of understanding the problems with such an approach is
as follows: The coupling of a scalar (dilaton) field $\P (X)$ to a
massive particle will have the form 
\beq S_I = m\int d\tau
e^\P\sqrt{-\dot{X}^\mu \dot{X}^\nu \eta_{\mu \nu}}, 
\eeq{mcoup} 
where
$\eta_{\mu \nu}$ is the Minkowski metric and dot denotes
$\tau$-derivative. This coupling may be interpreted as introducing
either an $X$-dependent mass, or the trace of the metric as an
independent degree of freedom. To be able to study massless particles
as well, we rewrite (\ref{mcoup}) on a first order form 
\beq
\tilde{S_I}=-\half\int d\tau\left( g^{-1}e^{2\P}\dot{X}^2-gm^2\right),
\eeq{mfo} 
where $g$ is an auxiliary ``einbein''. In the limit $m\to
0$, the field redefinitions 
\beq \hat g \equiv ge^{-2\P}, \qquad \hat
\P \equiv \P, 
\eeq{red} 
remove the coupling between the particle and
the dilaton field, thus revealing the problem with interaction of a
scalar field with a massless particle. From a different point of view
this just reflects the conformal invariance of the action for massless
particles. For $m = 0$, we may think of the coupling in (\ref{mfo}) as
resulting from a conformal transformation of the $X^\mu$'s and the
redefinition (\ref{red}) as the corresponding conformal transformation
of $g$.

The tensionless string, as discussed in e.g
\cite{Ulf}-\cite{lizz}, is the analogue of a massless particle in
one dimension higher. In fact, in a particular gauge, a tensionless
string just describes a set of massless particles moving subject to a
constraint. It is thus not surprising that the abovementioned
difficulty in coupling a scalar field extends to the tensionless
string too. In what follows we investigate the possible (classical)
interactions for tensionless strings using the ``action at a
distance'' formalism, pioneered for particles in \cite{Feynman} and
extended to strings in \cite{Kalb}. We find that tensionless strings
couple to symmetric and antisymmetric second rank tensor fields, but
not to dilatons.

\section{Interactions}

In this section we want to discuss the possible interactions of tensionless
strings via space-time fields. To begin with we give a brief summary of the
corresponding treatment of the tensile string, as presented in \cite{Kalb}.
The method used is that of
the direct interstring action formalism
(see e.g.\cite{Feynman} and references therein).

Let $X_a^\mu$ denote the coordinates of string number $a$ in the
$D$-dimensional string target space. The corresponding string world
sheet coordinates are $\{\xi^i_a\}\equiv \{\tau_a,\s_a\}$.  We write
the total action $S$ for a collection of interacting tensile strings
as $S=S_F+S_I$.  The free string action $S_F$ is the sum of the free
actions for each individual string, $S_F=\sum_a S_{aF}$, where
$S_{aF}$ is written in Nambu-Goto form 
\beq S_{aF}=T\int d^2\xi_a
\sqrt{-\sigma_a^{\mu\nu}\sigma_{a\mu\nu}}, \eeq{intt} 
where 
\beq
\s_a^{\mu\nu}\is\e^{ij}\d_i X^\mu\d_j X^\nu=\dot{X}_a^\mu X_a'^{\nu}-
X_a'^{\mu}\dot{X}_a^\nu , 
\eeq{dsig} 
dot represents $\tau_a$
derivative and prime represents $\s_a$ derivative.  (The expression
under the square root in (\ref{intt}) is indeed minus the determinant
of the induced metric.)

The interaction part $S_I$ is given by\footnote{We will write all sums
over strings explicitly as $\Sigma$.}  \cite{Kalb} 
\beq
S_I=\sum_{a<b}\int \int d^2\xi_ad^2\xi_b
R_{ab}({\sl{g}};X_{a,b};\d_i{X}_{a,b})\, .  
\eeq{inntt} 
It is assumed
that closed as well as open strings contribute to the sums.  The
functions $R_{ab}=R_{ba}$ depend on the string coordinates and their
derivatives, and ${\sl{g}}$ is the string coupling constant. We will
assume that the dependence on the derivatives is through
$\sigma_{\mu\nu}$, in analogy to the particle case. (The tangent to
the world line being replaced by the tangent bi-vector to the world
sheet.)

The $X^\mu$-equations of motion found by varying $S$ are
\beq
2TD_a^\nu(\frac{\sigma_{a\mu\nu}}{\sqrt{-\sigma_a^2}})=
\sum_{b\neq a}\int d^2\xi_b\Delta_{b\mu} R_{ab}\, ,
\eeq{equ}
where we use 
\ber
&&\sigma_a^2\equiv \sigma_a^{\mu\nu}\sigma_{a\mu\nu}\, ,\cr
&&\cr
&&D_a^\mu\equiv
\e^{ij}\d_jX^\mu_a\d_i\, ,\cr
&&\cr
&&\Delta_{a\mu}\equiv \frac{\partial}{\partial
X_a^\mu}-\frac{\d^2}{\d \xi_a^i\d (\d_iX_a^\mu )}\, .
\eer{eqnd}
It can be shown from (\ref{equ}) that the $R_{ab}$'s are invariant under 
separate
reparametrizations of the world sheets,
\cite{Kalb}. For open strings (\ref{equ}) has to be supplemented by boundary 
conditions
at
$\s_a = 0,\pi$:
\beq
2T\frac{\sigma^\mu_{a\nu}}{\sqrt{-\sigma_a^2}}\dot{X}^{o,\pi}_{a\mu} =
-\sum_{b\neq a}\int d^2\xi_b\frac
{\partial R_{ab}}{\partial X_b'^{\nu}}\, .
\eeq{bdy}

The expressions for $R_{ab}$ representing graviton, anti-symmetric tensor and 
dilaton 
exchange
between tensile strings were found in
\cite{Kalb} to be
\ber
R^G_{ab}&=&{\sl{g}}^2\frac{\sigma^{\mu\nu}_a\sigma_{a\nu}^\rho}{\sqrt{-\sigma_a^2}}
\frac{\sigma_{b\mu\alpha}\sigma_{b\rho}^\alpha}{\sqrt{-\sigma_b^2}}G\, ,\cr
&& \cr
R^B_{ab}&=&{\sl{g}}^2\sigma_a^{\mu\nu}\sigma_{b\mu\nu}G\, ,\cr
&& \cr
R^\P_{ab}&=&{\sl{g}}^2\sqrt{-\sigma_a^2}\sqrt{-\sigma_b^2}G\, ,
\eer{slns}
where
$G\is G(s^2_{ab})$ represents a Green's function of the appropriate kind and
$s^2_{ab}\is (X_a-X_b)^2$.

We now want to repeat the analysis for (bosonic) tensionless strings. There 
is no 
formulation of
these strings corresponding to the Nambu-Goto action used above, i.e., 
without an 
auxiliary field.
The closest we can get is to mimic equation (\ref{mfo}) for the particle, 
(without the
$\P$-field), and write the tensile string action with an auxiliary field 
\cite{Ulf},
\beq
\tilde{S_a}=-\half\int d^2\xi_a\left( g^{-1}_a\s_a^2-g_aT^2\right),
\eeq{Tfo}
which leads to the $T\to 0$ action
\beq
S^0_{aF}=-\half\int d^2\xi_a\left( g^{-1}_a\s_a^2\right).
\eeq{0fo}
Here the auxiliary field $g_a$ is a scalar density to ensure the $2D$ 
diffeomorphism
invariance of the action. We again want to consider a total action of the form 
$S=S_F+S_I$, with
$S_F$ being the sum of free string actions and $S_I$ representing the 
interactions 
between 
them.
This combination should represent a limit of the tensionful expression. We 
therefore assume the
same form for the interaction terms, i.e., with $R_{ab}$ being diffeomorphism 
invariant and
constructed from $\s_{\mu\nu}$'s. (The $R_{ab}$'s were already independent of 
$T$).
We will keep
a dependence on a, as yet unknown\footnote{The coupling constant for the 
tensile 
string has
an expression in terms of the dilaton expectation value, and counts the genus 
of the 
Riemann
surface describing the interaction. No such interpretation of a 
tensionless string coupling constant exists. We introduce it here by analogy 
to the tensile case.}, tensionless string coupling 
constant
${\sl{g}}'$. To ensure invariance of $R_{ab}$ we might contemplate a
dependence on
$g_a$. The corresponding possibility for the tensile string would be to
include a dependence on the auxiliary metric $g_{ij}$. This possibility is not 
considered in
\cite{Kalb}, but, again, there the string action is the Nambu-Goto action 
which is 
diffeomorphism
invariant by itself. In case we include a dependence on $g_a$, its
field equation is
\beq
g_a ^{-2}\sigma_a^2 +\frac{\partial}{\partial g_a}\sum_{b\neq a}
\int d^2\xi_bR_{ab}=0\, .
\eeq{geq}
From this we see that such a dependence will take us outside the class of 
tensionless 
strings,
(where $\s^2_a=0$), in general. For the relation (\ref{geq}) to imply 
$\s^2_a=0$, the
second term must be proportional to the first. But this means that $R_{ab} 
\propto
\s^2_a\s^2_b$, which is unacceptable, since then it will vanish on-shell. 
We will
hence assume that $R_{ab}$ is independent of the $g_a$'s.

Under the above assumptions, the
equations of motion that follow from the
$X^\mu$-variation of the total action
$S$ are
\beq
D^\nu _a(g^{-1}_a\sigma_{a\mu\nu})=\sum_{b\neq a}
\int d^2\xi_b\Delta_{b\mu}R_{ab}\, .
\eeq{eqS}
It serves as a gratifying check 
that the
reparametrization invariance of $R_{ab}({\sl{g}}',X,\d X)$, (no dependence 
on $g_a$),
follows from (\ref{eqS}) when $\s^2_a=0$.
In addition to (\ref{eqS}) we also find the boundary conditions for open 
tensionless strings
\beq
g_a^{-1}\sigma^\mu _{a\nu}\dot{X}^{0,\pi}_{a\mu}=
\sum_{b\neq a}\int d^2\xi_b\frac{\partial R_{ab}}{\partial {X'}_b ^\nu}\, .
\eeq{bdy0}
Multiplying both sides of eqn. (\ref{bdy0}) by ${X'}_a^{\nu 0,\pi}$ we find a
constraint on the $R_{ab}$'s, namely that
\beq
{X'}_{a\nu}^{0,\pi}\sum_{b\neq a}\int d^2\xi_b\frac{\partial R_{ab}}{\partial 
{X'}_b ^\nu}=0\, .
\eeq{conR}
We deal with (\ref{bdy0}) and (\ref{conR}) as follows: In an orthonormal 
gauge where $X'\dot X=0$, the l.h.s. of (\ref{bdy0}) is proportional to 
${\dot X}^2X'$. We take as a boundary condition on open tensionless strings 
$X'=0$ in this gauge. This requires $R_{ab}$ to vanish on the boundaries too, 
by (\ref{bdy0}). This is satisfied for the $R_{ab}$'s constructed below.
The equations (\ref{eqS}) and (\ref{bdy0}) determine the motion, and the 
dynamics of tensionless strings when the
interaction terms $R_{ab}$ are specified. We will next turn our attention 
to the
explicit forms of these functions.

The expression for graviton exchange given in (\ref{slns}) for the
tensile case is of the form $\sim T_a^{\mu\rho}T_{b\mu\rho}$, where
$T_a^{\mu\nu}$ is the {\em space-time} energy-momentum tensor
corresponding to $S_{aF}$ (evaluated on the world sheet).
In the present case the space-time energy-momentum tensor is
\beq
\tilde{T}_a^{\mu\nu}\propto
g^{-1}_a\s^{\mu\g}\s^\nu_\g=g_a^{-1}\e^{ij}\e^{kl}\g_{jl}\d_iX^\mu\d_kX^\nu\, ,
\eeq{tmunu}
(with $\g_{ij}$ the induced metric). Since the $R_{ab}$'s should be 
independent of
$g_a$, we try to construct a tensorial $R_{ab}$ from 
$g_a\tilde{T}_a^{\mu\nu}$. An
analogy to the tensile relation (\ref{slns}) would be to choose
\beq
R^G_{ab} = {\sl{g}}'^2\tilde{T}_a^{\mu\nu}\tilde{T}_{b\mu\nu}g_ag_bG
\eeq{g0}
as a candidate for the gravitational string-string interaction. 
(Here $G(s^2_{ab})$ is
again an appropriate Greens function.) This choice
has the wrong mass dimension\footnote{The mass dimension of {\sl{g}}' is 
taken to be 1, 
as for the
tensile string.} and is not diffeomorphism invariant, however. Instead we 
construct the
following expression
\beq
R^G_{ab}={\sl{g}'} ^2\sqrt{\s_a^{\mu\nu}\s^\rho _{a\nu}\s_{b\mu\alpha}\s^\alpha
_{b\rho}}G\, .
\eeq{gg0}
This expression {\em is} reparametrization invariant, has the tensor structure 
appropriate for
graviton-like interaction and is of the right dimension.

The expression for exchange in the Kalb-Ramond
sector can be copied directly from eq.(\ref{slns}) without any changes
\beq
R_{ab}^B={\sl{g}'} ^2\s_a^{\mu\nu}\s_{b\mu\nu}G\, .
\eeq{bmunu}
This expression is reparametrization invariant (and non-zero) also in the 
present case.

Finally we consider possible scalar interactions. The dilaton interaction in
(\ref{slns}) is proportional to the products of the square roots of the 
traced energy-momentum tensors of strings $a$ and $b$. These are the only
scalars that we can form from our $\s$-building blocks of the right 
dimension. However,
in our case these traces are proportional to $\s^2_a$ and thus vanish, a 
sign of the
space-time conformal invariance of the tensionless strings. We conclude that 
dilaton interactions are absent.\footnote{We note that a decoupling of the 
dilaton from supersymmetric non critical strings, under certain circumstances, 
has  been discussed in the context of
tensionless strings in $M$-theory
\cite{Seiberg}.}

\section{Conclusions}

Our results are that tensionless strings may couple to gravitons and 
antisymmetric
tensor fields. This is in good agreement with the possible background 
geometry terms
that we may write down for a tensionless string. Using an equivalent 
formulation of
the tensionless string \cite{ils}
\beq
S^0=\int d^2\xi V^iV^j\d X^\mu \d X^\nu\eta_{\mu\nu}\, ,
\eeq{VV}
where the $V^i$'s are $2D$ vector density fields,
we see that the corresponding $\s$-model action is
\beq
\tilde{S}^0=\int d^2\xi \left( V^iV^j\d X^\mu \d X^\nu G_{\mu\nu}(X) 
+ \e^{ij}\d X^\mu \d X^\nu B_{\mu\nu}(X)\right)\, .
\eeq{VVS}
Here $G_{\mu\nu}$ and $B_{\mu\nu}$ are the background graviton and 
antisymmetric
tensor fields, respectively. Now, the dilaton is related to the trace of 
$G_{\mu\nu}$
and would couple as $e^\P\eta_{\mu\nu}$, just as described for the particle 
in the
introduction. Again, as for the massless particle, (\ref{VV}) has space time 
conformal
invariance\footnote{This can be extended to invariance of (\ref{VVS}) under 
conformal 
isometries.
} and such a coupling may be reabsorbed via a conformal 
transformation of
$V^i$. In the $\s$-model corresponding to (\ref{VVS}) for the tensile case, the
dilaton enters in two ways: Through a coupling to the $2D$-curvature scalar 
$R^2$ and
through an ambiguity in $G_{\mu\nu}$, 
(string metric vs Einstein metric, e.g.). The
first coupling is not available to us since we do not have a $2D$ metric and
representing the Euler characteristic in terms of (an integral of) $V^i$'s 
seems impossible. 
In any case
it arises as a one-loop effect, and our discussion is purely classical. 
The second
coupling was discussed above.

Couplings in the $\s$-model action for tensionless strings have previously been
discussed in \cite{zelt}, where space-time was restricted to four
dimensions.  There it was found that (in a special gauge) the $B_{\mu
\nu}$ field could be eliminated by a world sheet
reparametrization. This result crucially depends on the space time
dimension, and cannot be true in general.

\section{Acknowledgements}

We have benefitted from discussions with M.Ro\v cek, P.Saltsidis and 
B.Sundborg.
B.J. thanks ITP,
University of Stockholm for hospitality. 
B.J. also acknowleges partial support from NFR under grant no 110945/432. U.L.
acknowleges partial support from NFR under grant no 4038-312 and from 
NorFA under grant
no 96.55.030-O.

\end{document}